\documentclass[12pt]{article}
\usepackage[dvips]{epsfig}

\begin{document}
\begin{flushright}
OHSTPY--HEP--TH--98--005 \\
hep-th/9803170
\end{flushright}
\vspace{20mm}
\begin{center}
{\LARGE Non-Perturbative Spectrum 
   of Two Dimensional
 $(1,1)$ Super Yang-Mills 
 at Finite and Large $N$} \\
\vspace{20mm}
{\bf F.Antonuccio, O.Lunin, S.Pinsky} \\
\vspace{10mm}
{\em Department of Physics,\\ The Ohio State University,\\ Columbus,
OH 43210} 
\end{center}
\vspace{20mm}
\begin{abstract}
We consider the dimensional reduction of ${\cal N} = 1$
$\mbox{SYM}_{2+1}$ to $1+1$ dimensions, which has
$(1,1)$ supersymmetry. The gauge groups
we consider are U($N$) and SU($N$), where $N$ is a finite
variable.
We implement Discrete Light-Cone Quantization 
to determine non-perturbatively the  
bound states in this theory.
A careful analysis of the spectrum is performed at
various values of $N$, including the case where $N$
is large (but finite), allowing a precise measurement of the  
$1/N$ effects in the quantum theory. The low energy 
sector of the theory is shown to be dominated by string-like states. 
The techniques
developed here may be applied to any two dimensional
field theory with or without supersymmetry.

\end{abstract}
\newpage

\baselineskip .25in

\section{Introduction}

Solving for the non-perturbative properties of 
quantum field theories -- such as QCD -- is typically
an intractable problem. In order to gain some  
insights, however, a number of lower
dimensional models have been proposed as useful
laboratories in which to study QCD related
phenomena (for a review see\cite{bpp98}). 

In recent times, the role of low dimensional quantum
field theories has shifted rather dramatically following
the remarkable developments in string/$M$ theory.
The present literature on this subject is immense, but
a common theme appears to be emerging: there is 
more interesting physics in 
Yang-Mills theory than was
once thought reasonably possible. Besides the M(atrix) Model conjecture,
which formulates $M$ theory in terms of 
supersymmetric
quantum mechanics 
\cite{bfss97}, 
there is also a proposal by Maldacena \cite{maldacena} that 
large $N$ super Yang-Mills theories in various dimensions
are related to certain supergravity solutions.

All of these developments suggest that it would be
desirable to have a better understanding
of the non-perturbative 
properties of super Yang-Mills theory at
large (but finite) $N$, and in any dimension.   
Towards this end, we choose to study in detail
the bound state structure and spectrum of a two
dimensional field theory, which may be
obtained by dimensionally reducing $2+1$ dimensional
${\cal N} = 1$ super Yang Mills. 
Such a theory has already been investigated in the $N=\infty$
(or planar) approximation \cite{mss95}, and  is
believed to exhibit
the property of screening \cite{gkm96,ars97}. 
In this work  
we will allow the number of gauge colors, $N$, to be a finite variable.
This means we will be able
to monitor the behavior of the spectrum as $N$ is varied and made
arbitrarily large. Special attention is given to measuring
the precise effects on the spectrum due to $1/N$ contributions
in the quantum theory. 

Although we focus on one particular model in this paper, the 
techniques we develop here are applicable to {\em any} 
two dimensional field theory, with or without supersymmetry.

The organization of the paper may be summarized as follows;
in Section \ref{2Dsym}, we discuss the relevant features
of a (1,1) super Yang-Mills theory in $1+1$ dimensions, 
giving explicit
expressions for the (quantized) light-cone supercharges
formulated in the light-cone gauge.
Formulation of the DLCQ bound-state problem of this theory
is the subject of Section \ref{dlcq}, followed by a detailed
analysis of the corresponding numerical bound-state solutions in
Section \ref{numerical}.
In Section \ref{conclusions}, we conclude
with a perspective on future applications
of non-perturbative finite $N$ calculations
for arbitrary (super) Yang-Mills theories.

\section{$(1,1)$ Super Yang-Mills in $1+1$ Dimensions}
\label{2Dsym}
The theory we wish to study is readily obtained by
dimensionally reducing ${\cal N}=1$ $D=3$ super Yang-Mills
to $1+1$ dimensions. The resulting theory has $(1,1)$ supersymmetry,
and can be formulated in the light-cone frame. 
The details of this light-cone formulation  
appears in \cite{mss95}, to which we refer the reader for
explicit derivations. We simply note here that
the light-cone Hamiltonian $P^-$ is given in terms of the 
supercharge $Q^-$ via the supersymmetry
relation $\{Q^-,Q^-\} = 2 \sqrt{2} P^-$, where
\begin{equation}
    Q^-  =  2^{3/4} g \int dx^- \mbox{tr} \left\{
         ({\rm i}[\phi,\partial_- \phi ] + 2 \psi \psi ) \frac{1}{
    \partial_-} \psi \right\}. \label{qminus}
\end{equation}
In the above, $\phi_{ij} = \phi_{ij}(x^+,x^-)$ and 
$\psi_{ij} = \psi_{ij}(x^+,x^-)$
are $N \times N$ Hermitian matrix fields representing the physical
boson and fermion degrees of freedom (respectively) of the theory,
and are remnants of the physical transverse degrees of freedom
of the original $2+1$ dimensional theory. 
This is a special feature of light-cone quantization in light-cone 
gauge: all unphysical degrees of freedom present in the original
Lagrangian may be explicitly eliminated. There are no ghosts.

For completeness, we write the additional relation 
$\{Q^+,Q^+\} = 2 \sqrt{2} P^+$ for the light-cone momentum $P^+$,
where 
\begin{equation}
    Q^+  =  2^{1/4} \int dx^- \mbox{tr} \left[
         (\partial_- \phi)^2 + {\rm i}  \psi \partial_- \psi   \right].
\end{equation}
The $(1,1)$ supersymmetry of the model follows from the fact 
 $\{Q^+,Q^-\} = 0$.
%
%
%
%
%
%
In order to quantize $\phi$ and $\psi$ on the light-cone, we
first introduce the following
expansions at fixed light-cone time $x^{+}=0$:
\begin{eqnarray}
\phi_{ij}(x^-,0)=\frac{1}{\sqrt{2\pi}}\int_0^{\infty}
\frac{dk^+}{\sqrt{2k^+}}\left(a_{ij}(k^+) e^{-ik^+ x^-}+a^\dagger_{ji}(k^+)
e^{ik^+ x^-}\right); \label{phiexp}\\
\psi_{ij}(x^-,0)=\frac{1}{2\sqrt{\pi}}\int_0^{\infty}
dk^+ \left(b_{ij}(k^+) e^{-ik^+ x^-}+b^\dagger_{ji}(k^+)
e^{ik^+ x^-}\right). \label{psiexp}
\end{eqnarray}
We then specify the commutation relations
\begin{equation}
\left[a_{ij}(p^+ ),a^\dagger_{lk}(q^+)\right]=\left\{ b_{ij}(p^+ ),
b^\dagger_{lk}(q^+)\right\}=\delta (p^+ -q^+)\delta_{il}\delta_{jk}
\label{uncomm} 
\end{equation}
for the gauge group U($N$), or
\begin{equation}
\left[a_{ij}(p^+ ),a^\dagger_{lk}(q^+)\right]=\left\{ b_{ij}(p^+ ),
b^\dagger_{lk}(q^+)\right\}=\delta (p^+ -q^+)\left(\delta_{il}\delta_{jk}-
\frac{1}{N}\delta_{ij}\delta_{kl}\right)
\label{suncomm}
\end{equation}
for the gauge group SU($N$)\footnote{We assume the normalization 
${\mbox tr}[T^a T^b] = \delta^{ab}$, where the $T^a$'s are the 
generators of the Lie algebra of SU($N$).}.  

For the bound state eigen-problem
$2P^+ P^- |\Psi> = M^2 |\Psi>$, we may restrict to the
subspace of states with fixed light-cone momentum $P^+$, 
on which $P^+$ is diagonal, and so the bound state problem is 
reduced to the diagonalization
of the light-cone Hamiltonian $P^-$.
Since $P^-$ is proportional to the square of the supercharge $Q^-$,
any eigenstate $|\Psi>$ of $P^-$ with mass squared $M^2$ gives
rise to a natural  degeneracy in the spectrum because
of the supersymmetry algebra---all four states below have the same mass: 
\begin{equation}
    |\Psi>, \hspace{4mm} Q^+ |\Psi>,\hspace{4mm}  Q^- |\Psi>, 
\hspace{4mm}  Q^+ Q^- |\Psi>.  
\end{equation}
Although this  degeneracy is realized in the continuum
formulation of the theory, this property will not
necessarily survive if we choose to discretize the 
theory in an arbitrary manner. However, a nice  
feature of DLCQ is that it does
preserve the supersymmetry (and hence the {\it exact} four-fold
degeneracy) for any resolution. In the context of numerical 
calculations, this reduces (by a factor of four) 
the size of the DLCQ matrix that
needs to be diagonalized. 

The explicit expression for $Q^-$,
in the momentum representation 
is now obtained by substituting 
the quantized field expressions (\ref{phiexp}) and (\ref{psiexp})
directly into  the definition of the supercharge (\ref{qminus}).
The result is:
\begin{eqnarray}
\label{Qminus}
Q^-&=& {{\rm i} 2^{-1/4} g \over \sqrt{\pi}}\int_0^\infty dk_1dk_2dk_3
\delta(k_1+k_2-k_3) \left\{ \frac{}{} \right.\nonumber\\
&&{1 \over 2\sqrt{k_1 k_2}} {k_2-k_1 \over k_3}
[a_{ik}^\dagger(k_1) a_{kj}^\dagger(k_2) b_{ij}(k_3)
-b_{ij}^\dagger(k_3)a_{ik}(k_1) a_{kj}(k_2) ]\nonumber\\
&&{1 \over 2\sqrt{k_1 k_3}} {k_1+k_3 \over k_2}
[a_{ik}^\dagger(k_3) a_{kj}(k_1) b_{ij}(k_2)
-a_{ik}^\dagger(k_1) b_{kj}^\dagger(k_2)a_{ij}(k_3) ]\nonumber\\
&&{1 \over 2\sqrt{k_2 k_3}} {k_2+k_3 \over k_1}
[b_{ik}^\dagger(k_1) a_{kj}^\dagger(k_2) a_{ij}(k_3)
-a_{ij}^\dagger(k_3)b_{ik}(k_1) a_{kj}(k_2) ]\nonumber\\
&& ({ 1\over k_1}+{1 \over k_2}-{1\over k_3})
[b_{ik}^\dagger(k_1) b_{kj}^\dagger(k_2) b_{ij}(k_3)
+b_{ij}^\dagger(k_3) b_{ik}(k_1) b_{kj}(k_2)]  \left. \frac{}{}\right\}.
\end{eqnarray}
 
In ordinary DLCQ calculations, one chooses to
discretize the light-cone Hamiltonian $P^-$. However 
it was pointed out in \cite{mss95} that supersymmetric
theories admit a natural DLCQ formulation in terms of
discretized supercharges. This ensures that {\em supersymmetry
is preserved even in the discretized theory}. 
Before proceeding with the DLCQ formulation of the bound state
problem, we note that 
 for the gauge group U($N$), massless states
appear automatically because of the decoupling of the U($1$)
and SU($N$) degrees of freedom that constitute U($N$). More explicitly,
we may introduce the U(1) operators
\begin{equation}
         \alpha (k^+)  =  \frac{1}{N}\mbox{tr} [a(k^+)] 
\hspace{4mm} \mbox{and} 
   \hspace{4mm} \beta (k^+)  =  \frac{1}{N}\mbox{tr} [b(k^+)],
\end{equation}
which allow us to decompose any U($N$) operator into a sum of
U(1) and SU($N$) operators:
\begin{equation}
            a(k^+) = \alpha (k^+)\cdot \mbox{${\bf 1}_{N\times N}$} + 
{\tilde a}(k^+)
 \hspace{4mm} \mbox{and} \hspace{4mm}
      b(k^+) = \beta (k^+)\cdot \mbox{${\bf 1}_{N\times N}$}
 + {\tilde b}(k^+), 
\end{equation}
where ${\tilde a}(k^+)$ and ${\tilde b}(k^+)$ are traceless $N \times N$
matrices. If we now substitute the operators above into the
expression for the supercharge (\ref{Qminus}), 
we find that
all terms involving the U(1) factors $\alpha(k^+), \beta(k^+)$ vanish -- only
the SU($N$) operators ${\tilde a}(k^+),{\tilde b}(k^+)$ survive.
i.e. starting with the definition of the U($N$)
supercharge, we end up with the definition of the SU($N$) supercharge.
In addition, the (anti)commutation relations
$[{\tilde a}_{ij}(k_1),\alpha^{\dagger}(k_2)] = 0$ and
$\{{\tilde b}_{ij}(k_1),\beta^{\dagger}(k_2)\} = 0$ imply 
that this supercharge acts only on the SU($N$) creation operators
of a Fock state - the U(1) creation operators only introduce
degeneracies in the SU($N$) spectrum. Clearly, since $Q^-$ has no
U(1) contribution, any Fock state made up of only U(1)
creation operators must have zero mass. The non-trivial
problem is therefore solving for
SU($N$) bound states.

\section{Discretized Light-Cone Quantization at Finite $N$}
\label{dlcq}

In order to implement the DLCQ formulation \cite{dlcqpapers}
of the theory, we simply restrict the 
momenta $k_1,k_2$ and $k_3$ appearing in equation (\ref{Qminus}) to
the following set of allowed momenta: $\{\frac{P^+}{K},\frac{2P^+}{K},
\frac{3P^+}{K},\dots \}$. Note that
we omit the zero momentum modes \cite{pin97a,mrp97},
which are not expected to affect the massive spectrum.
Here, $K$ is some arbitrary positive integer,
and must be sent to infinity if we wish to recover the continuum 
formulation of the theory. The integer $K$ 
is called the {\em harmonic resolution},
and $1/K$ measures the coarseness of our discretization\footnote{Recently,
Susskind has proposed a  connection between the 
harmonic resolution arising from the DLCQ of $M$ theory, and the
integer $N$ appearing in the U($N$) gauge group for M(atrix) Theory
(namely, they are the same) \cite{suss97}.}.
Physically, $1/K$ represents the smallest unit of longitudinal
momentum fraction allowed for each parton. 
As soon as we implement the DLCQ procedure, which is 
specified unambiguously
by the harmonic resolution $K$, the integrals appearing
in the definition of $Q^-$ are replaced by finite sums,
and the eigen-equation $(Q^-)^2 |\Psi \rangle = 
\lambda |\Psi \rangle $ is reduced to a finite matrix
problem. For sufficiently small values of $K$ (in this case 
for $K \leq 4$)
this eigen-problem may be solved analytically.
For values $K \geq 5$, we may still compute
the DLCQ supercharge analytically as a function of $N$,
but the diagonalization procedure must be performed 
numerically.

The details of how to construct the DLCQ 
light-cone supercharges in the model studied here
appear in reference \cite{mss95}. A similar 
model was also studied using this approach in \cite{hak95}.
 The only 
modification we make here is that we allow the
number of gauge colours, $N$, to be a finite (algebraic)
variable. This complicates things considerably.
The reason is rather simple. In the $N=\infty$ 
formulation, all fockstates may be written
as a {\em single} trace of creation operators,
\begin{equation}\label{fockeg}
     |\Psi\rangle \sim \mbox{tr}[c^{\dagger}(k^+_1) \cdots
           c^{\dagger}(k_n^+)]|0\rangle
\end{equation}
($c^{\dagger}(k^+)$ represents either a boson or fermion carrying
longitudinal momentum $k^+$), since individual Fockstates
that involve a product of two or more traces couple to these 
states like $1/N$, and are therefore completely decoupled
in the limit $N=\infty$.
This gives rise to decoupled sectors that are characterized
by the number of traces appearing in each Fock state.
In addition, color interactions
in the light-cone Hamiltonian (or supercharge) simplify 
when $N=\infty$, since splitting or
joining interactions occur between {\em adjacent} 
color-contracted partons
 in a Fock state. This dramatically simplifies the 
representation of any light-cone operator on the Hilbert
space of single trace Fock states. This property 
also tremendously simplifies the evaluation of inner products.  
It is sometimes helpful
to think of a single trace state as a closed string
made up of `string bits' \cite{thorn}. Multiple-trace
states are therefore multi-string states, and
the string coupling is given by $1/N$. For $N=\infty$,
these multi-trace states are just free non-interacting
closed `strings'. Splitting and joining of these strings is only 
possible when $N$ is finite.

Of course, as soon as we allow $N$ to be finite, we have 
to give up all of these wonderful simplifications!
In computational terms,
this usually means that the most time consuming 
part of a DLCQ calculation is the evaluation inner products 
for many parton  Fock
states, which is relatively
trivial in the $N=\infty$ case. Of course, 
the processing time involved in
calculating the representation of the light-cone Hamiltonian
relative to the discretized Fock basis 
is augmented considerably due to these complications.

Nevertheless, we feel justified in dealing with these
complications, since a number of interesting physical
properties associated with the dynamics of super Yang-Mills theory
are expected to arise as `$1/N$ effects' in the 
quantum theory.\footnote{Maldacena has recently 
argued that the $1/N$ effects for
a particular class of super Yang-Mills theories account
for Hawking radiation in a corresponding
class of space-time geometries \cite{maldacena}.}

In practical terms, the complexities cited above 
simply restrict how large the harmonic resolution, $K$, is allowed
to be in numerical computations. In the present study,
we could manage only $K \leq 8$ (about 2000 states altogether
for $K=8$), and we expect that higher values of $K$ could be probed
if more powerful machines and more efficient code were available.\footnote{
Numerical calculations were performed using a desk-top PC, and 
the computer code was written for Mathematica Version 3.0.}  

Before proceeding to discuss our numerical results, we 
point out that one may significantly reduce the  
computational complexity of  setting up the DLCQ supercharge
by taking advantage of the simple fact that the U($N$) 
and SU($N$) supercharges are equivalent. 
From a computational point
of view, the commutation relations for U($N$) matrices (eqn \ref{uncomm})
are simpler than the  SU($N$) relations (eqn \ref{suncomm}),
and so it would be desirable to work with the U($N$) 
basis even when we are interested in solving
for  SU($N$) bound states.  
It turns out that if one constructs a 
basis of U($N$) Fock states, and then discards all states 
that contain a trace of a single parton, then 
the corresponding spectrum of the U($N$) theory
on this modified basis yields
the same spectrum as the SU($N$) theory. 
Of course,
constructing the U($N$) supercharge requires much less 
computational effort,
and we therefore employ this strategy when solving for 
SU($N$) bound states when $K$ is large. A more thorough
discussion of this technique will appear elsewhere \cite{alp98c}. 
Of course, the DLCQ program we use can do both SU($N$) and 
U($N$) independently, and the above procedure can be checked
explicitly for $K \leq 6$ (it works!). This method is expected
to play a crucial role when solving for  SU($N$) bound states in 
more complicated two dimensional theories.

\section{Numerical Bound State Solutions}
\label{numerical}
There are two parameters in the DLCQ formulation
of the theory; the harmonic resolution $K$,
and the number of gauge colors $N$. 
This dependence on $K$ is of course an artifact of
the light-cone compactification scheme, 
 $x^- = x^- + 2 \pi R$, and in practice it is
eliminated by extrapolating the results at
finite $K$ to the continuum limit $K = \infty$.
Reliable extrapolations require careful analysis of the theory
as $K$ is steadily increased.
From equation (\ref{qminus}), 
one sees that the two dimensional Yang-Mills 
coupling constant $g$ factors out in the definition of the light-cone
supercharge, and so the only adjustable coupling constant 
in the theory is the parameter $1/N$. This quantity
measures the 
strength of interactions between different trace sectors
in the Hilbert space,  where each sector
is characterized by the number of traces
in each Fock state. Of course, in the limit $N=\infty$,
these sectors are completely decoupled.
It is therefore an interesting physical problem
to investigate the behavior of the theory 
when $N$ is allowed to be finite and large.
Bound states will be a superposition of Fock states containing 
any number of traces, but interactions 
between different sectors will be weak.

\begin{figure}[ht]
\begin{center}
\epsfig{file=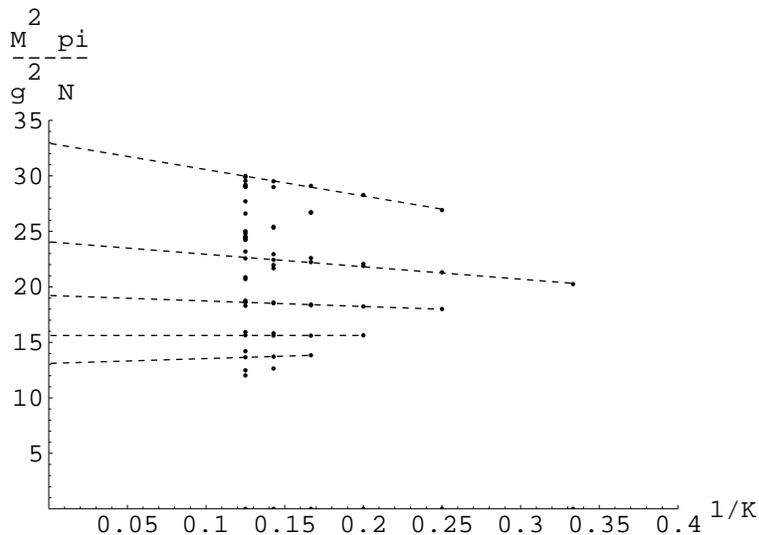}
\end{center}
\caption{{\small Bound State Masses $M^2$ (in units $g^2 N/\pi$)
versus $1/K$ for $N=10$.} 
\label{one}}
\end{figure}

Our numerical work involved
solving the DLCQ SU($N$) bound state equations for $2 \leq K \leq 8$,
and then extrapolating the results to obtain estimates for
continuum bound state masses. Supersymmetry
in the DLCQ formulation gives rise to an
obvious exact two-fold mass degeneracy between
bosons and fermions, but there is an 
additional two-fold degeneracy for each massive
boson and fermion bound state. We therefore have 
an exact four-fold degeneracy in the spectrum
of massive bound states. 
Figure \ref{one}
is a summary of the low mass spectrum 
(i.e. only eigenvalues less than 30 are plotted) 
that were obtained for $3 \leq K \leq 8$, and for $N=10$. 
The vertical axis measures the bound state mass squared 
$M^2 \equiv 2P^+P^-$, in units of $g^2 N/\pi$. 

At resolution $K=2$ there are precisely two massless SU($N$) bound
states (one boson and one fermion), each consisting of two
partons.
At $K=3$, massive bound states begin to appear in the spectrum,
and are four-fold degenerate (two bosons and two fermions). 
If these solutions signify the presence 
of true bound states in the continuum, 
one expects that their structure 
will persist as $K$ is increased. In order to check this,
one must look at the Fock state content of a bound state
at different resolutions, 
and see whether the same approximate structure  
is preserved as we increase $K$; whether the wave functions 
begin to converge or not is an indicator of whether the 
continuum bound state might 
be normalizable or not. 
Of course, as we continue to increase
$K$, new states will appear in the (discretized) Fock space
that are not related to any states at smaller resolution.
These will signify the onset of additional bound states
that may also be followed as $K$ is increased. The
first appearance of a state at a given resolution
may be thought of as a  ``trail head'' \cite{ghk97} of
the corresponding continuum state.
At large $N$,
this procedure of finding `trails of bound states'
is rather straightforward to carry out (although tedious).
For intermediate values -- say $N=10$ -- 
the Fock state content
of any bound state is complicated considerably due to the non-trivial
mixing between states with differing numbers of traces, and
the procedure of following the trail of a bound state at 
different resolutions must be administered with care. 

In Figure \ref{one}, we illustrate this procedure
for the case $N=10$, where the dashed
lines represent trails of particular
bound states, and we have extrapolated
these curves to estimate continuum ($K=\infty$)
bound  state masses (we move from right to left on these curves as
$K$ is increased). 
In general, as we increase
the resolution,
states pick up additional Fock state 
contributions with a larger number of partons,
but the approximate structure seen at lower resolutions is 
still clearly visible provided $N$ is large. In order
to determine the trails of these states for intermediate
values of $N$, say $N=10$, we first consider the trail 
of a state for large $N$ (say $N =100$ or $1000$), and then
tune the value of $N$ down to the desired smaller value; this effects
a smooth change in wave function amplitudes, but the type
of Fock states in the Fock state expansion remains unchanged.
One can therefore be confident that one is tracking the correct state.

There are a number of striking features
in the DLCQ spectrum of the theory. Firstly,
the low energy spectrum appears to be dominated by 
string-like states; each time we increase the resolution,
a new massive state appears in the spectrum which
is lighter than any of the massive states that
appeared at a lower resolution.
This can be seen in Figure \ref{one}.
In addition, the average number of partons in these states 
increases commensurately with the resolution $K$.
So in the continuum $K \rightarrow \infty$, one can expect
the existence of very light bound states that 
have an arbitrarily large number of partons. 
Since the spectrum is bounded from below (by supersymmetry),
one deduces the existence of an accumulation point in the 
mass spectrum, which we denote by $M_c$.
Bound states with masses at (or at least sufficiently near) this point 
are expected to behave like strings made
out of an essentially infinite number of string `bits'.
Evidently, pair creation of partons seems
to be energetically favored.
This behavior directly contrasts what is observed
in many other models studied in the same framework \cite{bpp98}; 
namely, 
the mass of a state generally increases with
the average number of partons in its Fock state expansion.

\begin{figure}[ht]
\begin{center}
\epsfig{file=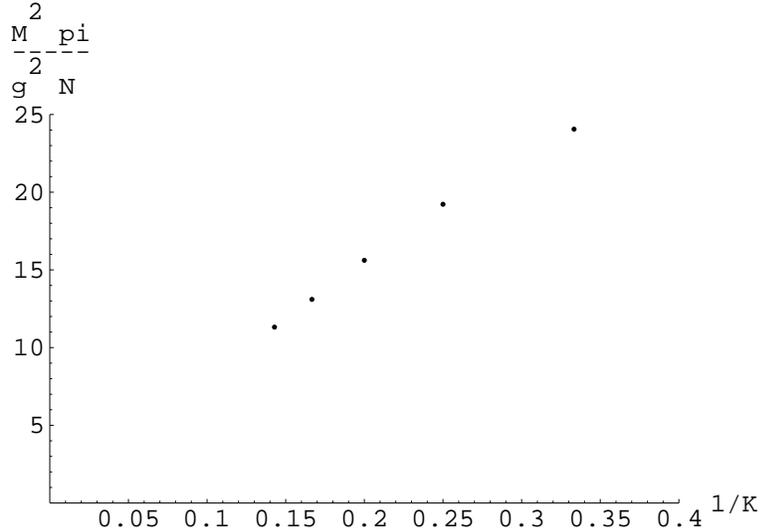}
\end{center}
\caption{{\small Extrapolated continuum masses of lightest massive 
bound states for different resolutions $K$,
and for $N=10$. The vertical axis is the extrapolated continuum
mass of the lightest non-zero mass eigenstates that
first appear at resolution $K$. The extrapolation of these
points to $K=\infty$ gives an estimate for the accumulation
point $M_c^2$ in the spectrum.} 
\label{two}}
\end{figure}

One interesting question that we are unable to answer
is whether $M_c$ is zero or not.
We know that massless states do exist
at any resolution (there are in fact $2^{(K-1)}$ of them
at the resolution $K$), and it might seem reasonable  
that these light massive states approach
the already existing massless states in the limit
$K \rightarrow \infty$ limit. 
We see from Figure \ref{two} that the prediction for $M_c$,
which is the extrapolation of the points to $K=\infty$, 
appears to be very close to zero, if not exactly zero.
The horizontal axis is specified by $1/K$, where $K$ is the 
resolution at which the lightest non-zero mass 
eigenstate first appears
(i.e. has a `trail head'), and the vertical axis is its
extrapolated continuum mass (i.e. where the extrapolation
curves in Figure \ref{one} intersect the vertical axis).
Due to extrapolation errors, and the low resolutions
that were attainable, the uncertainties 
in Figure \ref{two} are expected to be quite large. 
  
Another interesting feature of the DLCQ spectrum is that
the extrapolation curves (dashed lines in Figure \ref{one})
are relatively {\em flat}. This means that by performing a 
relatively trivial calculation at $K=3$ or $4$, one is able
to estimate the continuum bound state mass perhaps within 
ten percent of the actual continuum value (assuming no
pathologies in the DLCQ spectrum for extremely large $K$). Of course,
for the massless states, the curve {\em is} perfectly flat,
and so we obtain exact information about the continuum
spectrum (i.e. that there are massless states).
There is an additional curious property about these 
massless states that appear in the DLCQ spectrum. 
It was shown recently \cite{alp98} that 
any normalizable massless bound state
in this theory is a superposition
of an infinite number of Fock states. Of course, when one works
in the DLCQ formulation, the number of Fock states
is finite. In our numerical analysis, however, we 
observe that the states are exactly massless at {\em any} resolution;
increasing the resolution increases the complexity of
the Fock state expansions of these massless states,
but the masses are always precisely zero.
Some very special cancellations are evidently
responsible for protecting these massless states from
receiving corrections due to the change in resolution.
Note that this is suggestive of some kind of `duality';
namely, for $K$ small, the problem is relatively easy to solve,
and has a simple description in terms of a small
number of degrees of freedom, 
while for $K \rightarrow \infty$, the complexity 
of the DLCQ problem increases dramatically, and the precise
description of corresponding bound states is in terms
of many more degrees of freedom. Nevertheless, 
the masses of certain states are preserved. 
It would be interesting to understand this from another 
point of view.

\begin{figure}[ht]
\begin{center}
\epsfig{file=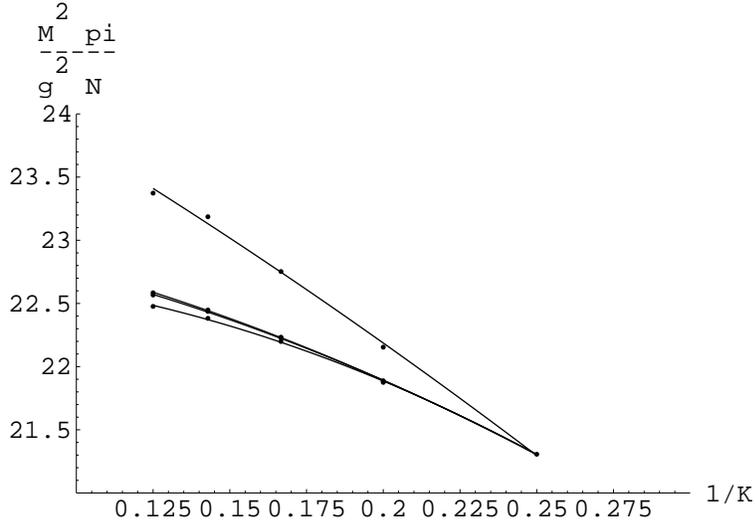}
\end{center}
\caption{{\small Bound state masses versus $1/K$ for 
different $N$; (a)$N=3$ (top curve), (b)$N=5$ (bottom curve),
         (c)$N=10$ (third from top) and (d)$N=100$
 (second from top). The $N=1000$ curve is
indistinguishable from the $N=100$ curve.} 
\label{three}}
\end{figure}

Since $N$ is an algebraic variable in our
calculations, we are able to investigate the
changes in the masses of states as $N$ is varied.
As an illustration, we consider the mass of a state
which has a ``trail head'' mass of $M^2=20.25 g^2 N/\pi$.
See Figure \ref{three}. The values of $N$ are $3$, $5$,
$10$, $100$ and $1000$, and we consider the range $3 \leq K \leq 8$
as usual.
 
Evidently, for $N=3$, the coupling $1/N$ is no longer
negligible, and there is an apparent shift in the estimated
continuum mass of the bound state. For $N >5$, convergence
to the large $N$ limit appears to be rather rapid.
The general behavior at large $N$  would be
consistent with the interpretation $\frac{M^2 \pi}{g^2 N} \sim
 a - \frac{b}{N^2}$, where $a$ and $b$ are positive
constants.  We do not have a term linear in $1/N$ since
one can show directly that $\frac{M^2 \pi}{g^2 N}$ is even under
the interchange\footnote{Recall that
the DLCQ Hamiltonian is an algebraic function of $N$,
and so we may analytically continue $N$ to non-integer 
or negative integer values by direct substitution!} $N \rightarrow -N$.
For $N=3$, it is clear from Figure \ref{three} that this picture
is no longer valid, and one would expect relevant
contributions at higher order in the $1/N$ expansion for $M^2$.

The multi-particle spectrum is a feature of the DLCQ spectrum that has only
recently
become of interest. It was pointed out in \cite{ghk97} that there
may be bound states in the DLCQ spectrum at resolution $K$
which may be thought of as  two non-interacting bound states;
this was verified by determining the masses $M^2(K-n)$ and
$M^2(n)$ of bound states
at resolutions $K-n$ and $n$ respectively 
($n$ is positive integral), and showing that the  
light-cone energy relation for two free particles,  
\begin{equation}
\label{twobodymass}
{ M^2 (K) \over K}={ M^2 (K-n) \over K-n}  + { M^2 (n) \over n},
\end{equation}
was obeyed. 
It is perhaps surprising the such a spectrum was found in \cite{ghk97},
since the
calculation was performed for $N=\infty$,
where the Hilbert space consists
of Fock states that are only {\em single} traces of parton
creation operators. There is no obvious way of
identifying single or multi-particle states in such a basis.
 In the case of finite but very large $N$, it
is easy to see how the spectrum approaches a many body continua;
the basis now consists of multi-trace states, but
interactions between bound states consisting of predominantly
single trace Fock states 
are suppressed by $1/N$. Two body continua in the spectrum
is therefore obvious in our finite analysis, and we will discuss this in more
detail shortly.
Nevertheless, it is tempting to speculate on a possible explanation 
for the presence of these ``multi-particle'' states in
the $N=\infty$ analysis; namely,
following the work \cite{alp98}, one  expects
even at very large (but finite) $N$ that any predominantly
two trace  bound state has a contribution from single trace
Fock states. One sees this directly in the DLCQ analysis,
of course. This suggests that it might be possible to ``see'' 
multi-trace bound  states in the $N=\infty$
spectrum (where there are no multi-trace states in
the Hilbert space) by virtue of the surviving
single trace contributions. A more thorough numerical investigation
will need to be carried out before this question
can be properly resolved, since the above argument rests heavily on the
dynamical properties of the theory.

\begin{figure}[ht]
\begin{center}
\epsfig{file=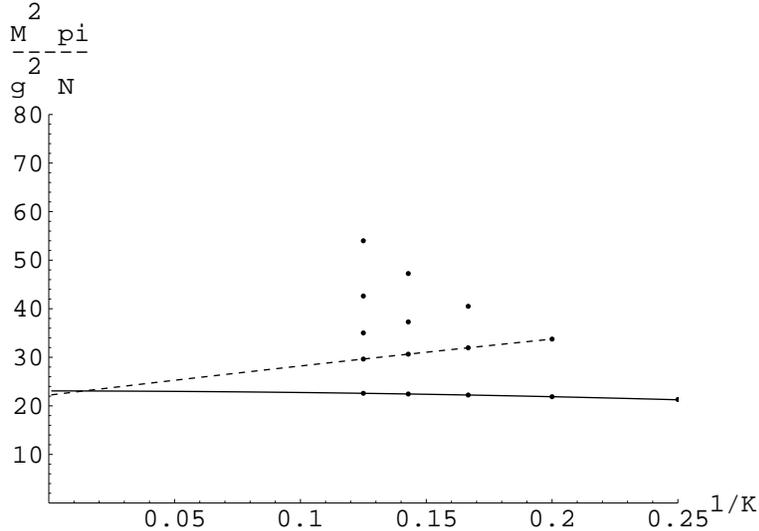}
\end{center}
\caption{{\small  Single and multi-particle bound state masses
 for $N=1000$. The solid line represents
the trail line of a predominantly single trace bound state.
The remaining points represent this  state  bound
with a massless bound state.
Equation (\ref{twobodymass}) predicts these masses
to high precision, and we therefore expect two body
continua in the spectrum in the limit $K \rightarrow \infty$.} 
\label{four}}
\end{figure}

We now return to the issue of multi-particle bound states
in the context of our finite $N$ calculations.
For very large values of $N$,
it is straightforward to identify in the bound state spectrum
those states that are essentially two loosely bound
particles; namely, any bound state that is predominantly
a superposition of two-trace Fock states are obvious
candidates. Of course, one needs to verify relation
(\ref{twobodymass}) before concluding that the bound state
admits such a `two free-particle' interpretation.
A representation of such calculations is given in 
Figure \ref{four}.

After solving the DLCQ bound state equations for different
resolutions, we are able to identify a predominantly
single trace bound state with three partons in the trace.
The extrapolated continuum mass is estimated by
the solid curve in Figure \ref{four}. One also discerns
many massless states consisting of two partons. At resolution $K=5$,
one finds a bound state consisting of two trace Fock states
that is readily identified as the two bound states mentioned
above that are essentially non-interacting.  Its mass
is predicted exactly by equation (\ref{twobodymass}). 
At resolution $K=6$, there are two ways to form the state and at 
resolutions
$K=7$ and $K=8$
there are three and four ways respectively to form the state
(see Figure \ref{four}). We find all
these states have a mass that 
is predicted by (\ref{twobodymass}) to
very high precision. What we are seeing is
therefore the discrete
realization of the two
body continuum spectrum. 
We have made a best fit to the lowest two-body mass at each
resolution,
and we see that the extrapolated value coincides
(within error) with the mass of the three parton
massive bound state. This is of course expected,  
since the mass at threshold of a massive and 
massless state is precisely the mass of the massive state.
In the DLCQ calculation, one finds that the masses of
these two body states are highly degenerate. The degeneracy
of the massive state is $4$, while the degeneracy of the massless states
is $2^{({\tilde K}-1)}$, where ${\tilde K}$ is the resolution
of the two parton massless states appearing in the
two-trace states. The total degeneracy is therefore
expected to be $2^{({\tilde K}+1)}$, which is indeed observed
in the spectrum. 
As we
increase the
resolution the density of points will increase and effectively fill the
continuum as a dense subset. What we have
presented is an illustrative example, and we
in fact find the 
same pattern for other combinations of states as well,
including three body spectra, which all
occur at the expected mass. 

\begin{figure}[ht]
\begin{center}
\epsfig{file=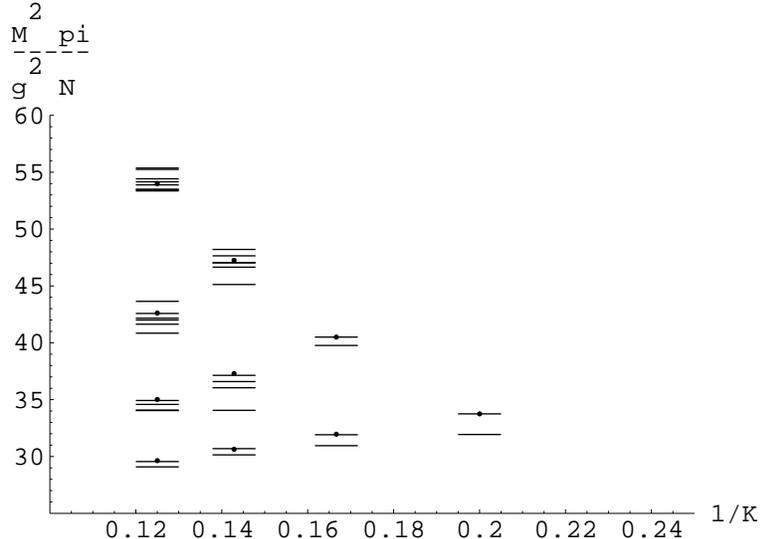}
\end{center}
\caption{{\small Mass splittings for multi-particle
bound states for $N=10$. The horizontal lines are bound state
masses, and the points are masses predicted 
by the two free particle formula (\ref{twobodymass}).
One sees the formation of bound states, suggesting that
for very large (but finite) $N$, the asymptotic degeneracy
of the spectrum could be quite complicated.} 
\label{five}}
\end{figure}

Of course, the above observations are expected as trivial 
realizations of the $1/N$ expansion.
What is of interest is the modifications in
the spectrum due to small but measurable contributions
if we allow $1/N$ interactions to become important.
As we mentioned earlier, studying the trail lines
(i.e. tracking a particular state at different resolutions
as in Figure \ref{one})
becomes increasingly difficult if $N$ is not very large.
To assist one in establishing  the correct
trail lines, it is helpful to first consider
identifying states at large $N$ (say N=1000),
and then following the state as $N$ is lowered to
the desired value. In Figure \ref{five} we perform this procedure,
starting with the two body spectrum 
represented in Figure \ref{four} (where $N=1000$), and 
then eventually arriving at the spectrum for $N=10$. 
The obvious difference between  Figure \ref{four} and 
Figure \ref{five} is that at $N=10$, the mass splittings in the
spectrum become discernible. Note that there is a discontinuous
change in the number of degrees of freedom at $N=\infty$,
since at this point, 
the bound states at large but finite $N$ will dissociate into
their constituent particles at $N=\infty$. 
The presence of multi-particle bound states for finite $N$
evidently provides scope for an
exponential growth in the density of states.

The points 
in Figure \ref{five} are the
values
predicted by equation (\ref{twobodymass}) at $N_c =10$. 
Most of the states are below the threshold indicating that at
$N_c =10$  the interaction that mixes the various trace sectors is
attractive and we
consider these states to be bound states. Some of the states are above the
threshold
implying that they are candidate continuum states.
The mass splittings introduced by $1/N$ 
interactions may push states above and/or below
threshold, depending on the details of
the interactions, and so determining the number of bound states 
in a theory
at finite $N$ is a highly non-trivial dynamical
question.

At this point, we remark that the additional interactions 
we introduce as a result of working with finite $N$ 
is suggestive of a system of weakly
interacting hadrons; the case $N=\infty$ is analogous
to a system of non-interacting colorless bound states,
while the $1/N$ effects introduce the many subtle interactions
that arise between colorless hadrons.

\section{Discussion}
\label{conclusions}

To summarize, we find that the low energy  
spectrum 
of $(1,1)$ SU($N$) super Yang-Mills in $1+1$ dimensions is 
dominated by string-like states. 
This followed from the observation that increasing
the DLCQ resolution introduces new lighter states
that have on average more partons in their Fock state 
expansion than states at smaller resolutions.
There is also strong numerical evidence that these states 
are normalizable, since one can keep track of these 
solutions as the resolution is increased, and we find that the Fock state 
amplitudes converge rapidly (See Figure \ref{one}).
It is therefore clear that pair creation of
partons in this theory is not energetically suppressed.
String-like states were also found in a theory involving
complex adjoint fermions, although their Fock state
content was much simpler \cite{anp97,pin97}
  
This immediately raises a question about the detailed 
structure of the spectrum. From supersymmetry,
masses are bounded from below, and we therefore infer
the existence of an accumulation point in the spectrum.
Near this point,  bound states  consist of   
an arbitrarily large number of partons. Whether
this accumulation point occurs at zero or positive mass was partly
addressed in Figure \ref{two}, and this still remains an open
question. Nevertheless, these results suggest that 
the fundamental degrees of freedom in the theory
(i.e. the normalizable bound states) may give rise to a continuous
spectrum starting at (or close to) zero mass. 
Whether this is the signature of an additional hidden dimension,
as was discussed in \cite{hak95} in the context of the non-critical
superstring in $2+1$ dimensions, or the manifestation
of screening \cite{gkm96,ars97}, is still unclear.
Nevertheless, it is clear that the model exhibits
remarkably complicated low energy dynamics.
 
One of the main goals of this work was to go beyond the
$N=\infty$ (or planar) approximation of gauge theories
in order to study $1/N$ effects (e.g see Figure \ref{three}). 
In the present context,
the quantity $1/N$ plays the role of a coupling constant,
and measures the strength of interactions between
sectors in the Hilbert space that are characterized
by the number of colorless traces in each Fock state.
For $N$ large (but finite), it is easy to identify
two `loosely bound' particles in the spectrum, since 
it will be made up
of predominantly two trace Fock states. We showed that
the same strategy adopted in \cite{ghk97} to
calculate the mass of two freely interacting 
bound states in the DLCQ spectrum applies equally well in 
the present context. Figure \ref{four} illustrates 
the manifestation of such `two body' continua in the DLCQ spectrum.    

For intermediate values of $N$, it is 
possible to measure the effects of $1/N$ interactions,
and we have presented an illustration of the 
mass splittings 
that occur in Figure \ref{five}.
It is a dynamical question whether an attractive force
will develop between particles that  
freely interact in the $N=\infty$ limit. Evidently, the formation
of bound states is favored in the present model, and
we are therefore faced with the interesting problem of counting the 
asymptotic degeneracies in the spectrum if $N$ is 
made arbitrarily large (but finite). We were unable to
address this question here. 
Note that the presence 
of very light string-like states suggests that the quantity $1/N$
plays the role of a string coupling constant \cite{thorn}.
It would be interesting to pursue these ideas further
in the context of a two dimensional super Yang-Mills
realization of the ten dimensional critical string \cite{dvv97}.

Finally,
it has become evident recently that the properties of
low dimensional super Yang-Mills may provide a non-perturbative
formulation of quantum theories with gravity. 
It would be interesting to explore this connection further
by performing the sort of non-perturbative analyses
presented here.

\medskip
\begin{large}
 {\bf Acknowledgments}
\end{large}

The authors would like to thank S.Tsujimaru
for his involvement at earlier stages in this work, and
I.Klebanov for helpful discussions.
 

\vfil

\end{document}